\documentclass[aps,reprint,twocolumn,amsmath,amssymb,superscriptaddress,
pra,
floatfix]{revtex4-2}
\usepackage{csquotes}
\usepackage{bm}%
\usepackage[colorlinks=true,linkcolor=blue]{hyperref}%
\expandafter\ifx\csname package@font\endcsname\relax\else
 \expandafter\expandafter
 \expandafter\usepackage
 \expandafter\expandafter
 \expandafter{\csname package@font\endcsname}%
\fi
\usepackage{bbm}
\usepackage{mathrsfs}
\usepackage{amsmath}
\usepackage{graphicx}
\addtocounter{MaxMatrixCols}{20}

\begin{document}

\title{Manipulation of diverse quantum correlations based on a hybrid optomagnomechanical system}%

\author{Xiaomin Liu}
\affiliation{College of Physics and Electronic Engineering, Shanxi University, Taiyuan 030006, China}
\affiliation{State Key Laboratory of Quantum Optics Technologies and Devices, Shanxi University, Taiyuan 030006, China}

\author{Rongguo Yang}
\affiliation{College of Physics and Electronic Engineering, Shanxi University, Taiyuan 030006, China}
\affiliation{State Key Laboratory of Quantum Optics Technologies and Devices, Shanxi University, Taiyuan 030006, China}
\affiliation{Collaborative Innovation Center of Extreme Optics, Shanxi University, Taiyuan 030006, China}

\author{Jing Zhang}
\email{zjj@sxu.edu.cn}
\affiliation{College of Physics and Electronic Engineering, Shanxi University, Taiyuan 030006, China}
\affiliation{State Key Laboratory of Quantum Optics Technologies and Devices, Shanxi University, Taiyuan 030006, China}
\affiliation{Collaborative Innovation Center of Extreme Optics, Shanxi University, Taiyuan 030006, China}

\author{Tiancai Zhang}
\affiliation{State Key Laboratory of Quantum Optics Technologies and Devices, Shanxi University, Taiyuan 030006, China}
\affiliation{Collaborative Innovation Center of Extreme Optics, Shanxi University, Taiyuan 030006, China}
\affiliation{Institute of Opto-Electronics, Shanxi University, Taiyuan 030006, China}

\date{\today}

\begin{abstract}
Flexible manipulation of quantum correlation resources enables the implementation of diverse quantum tasks based on hybrid quantum networks, where atom-magnon and optomagnonic entanglements and steerings play important roles. In this work, we propose an effective scheme to generate and manipulate quantum entanglements and steerings based on a hybrid optomagnomechanical system, which is composed of a polarizer, an optical cavity with YIG bridge as one end, and an atomic ensemble in it. According to the results of the parameter dependence of various quantum correlations, we can selectively generate bipartite and genuine tripartite entanglements and deterministically manipulate the concrete situation of bipartite, multipartite steerings, and collective pentapartite steering, by adjusting the polarization direction of the driving laser and the Tavis-Cummings coupling strength. Our all-optical controlled scheme is flexible, convenient, compact, and experimentally feasible, because multiple coupling channels can be tuned simultaneously. This work provides a new perspective for implementing specialized quantum tasks, such as hierarchical ultra-secure multi-user quantum communications. 
\end{abstract}

\maketitle



\section{Introduction}
The generation and manipulation of quantum correlations, especially quantum entanglement and steering, are necessary for various quantum tasks, including quantum teleportation~\cite{169}, quantum communication~\cite{965}, quantum networks~\cite{202266,2024579}, and quantum computation~\cite{602605}. 

The optomagnomechanical system is a promising candidate for generating diverse quantum correlations~\cite{015014,202206}, owing to its rich nonlinearity from the combination of cavity-optomechanics and magnomechanics. It exhibits distinct advantages for constructing hybrid quantum networks, as the magnons and phonons therein can couple to physical objects of arbitrary frequencies~\cite{093601,1098,281}. In general, a hybrid quantum network consists of distinct physical quantum systems, each of which can be chosen to implement dedicated quantum tasks with its inherent advantages: superconducting qubits excel in high-fidelity quantum information processing; optical signals are ideal for long-distance quantum communication; atomic ensembles offer reliable quantum information storage and read-out. Accordingly, the investigation of quantum correlations in hybrid optomagnomechanical systems incorporating atomic ensembles has attracted growing interest~\cite{023501,20242878,34998,s40507}. 
Moreover, the configuration of quantum resources in a quantum network should be flexible to facilitate the implementation of diverse quantum tasks and their dynamic switching. 

To effectively manipulate quantum entanglement and steering, various methods have been proposed, such as adjusting the coupling strength~\cite{024042}, pump proportion~\cite{042416,090501}, dissipation~\cite{052114}, or the relative phase between coupled modes~\cite{012421,123039}. For instance, the manipulation of bipartite steering between two magnons in a cavity magnonic system can be achieved by modulating the ratio of two cavity-magnon coupling strengths~\cite{024042}. That said, this method is somewhat limited in convenience and accuracy, as the positions of the two YIG spheres need to be refixed whenever the ratio is varied. Based on an optical parametric oscillator with a spatially structured pump, the manipulation of bipartite and multipartite steering of the down-converted Hermite–Gaussian modes can be obtained by varying the pump proportions via a spatial light modulator~\cite{042416}. Similarly, the modification of the entanglement structure of the modes generated from seven spatially multiplexed concurrent four-wave mixing processes driven by a spatially structured pump can be experimentally realized by tuning the pump power ratio~\cite{090501}. However, the precise generation and proportional modulation of distinct spatial modes are still experimentally challenging. Another approach involves introducing additional noise to experimentally realize the manipulation of the one-way steering direction in two-mode squeezed states~\cite{052114}, although this inevitably leads to a reduction in correlation and limits the secure communication distance. In addition, the manipulation of quantum entanglement and steering between two mechanical modes can be achieved by tuning the squeezing phase in a nonlinear whispering-gallery-mode optomechanical resonator~\cite{012421}. The optomechanical entanglements and steerings in a three-mode linked coupling system, where a membrane is placed in an optical cavity, can be selectively generated by adjusting the relative phase between the mechanical and cavity modes~\cite{123039}. Although the phase control method is effective, it can only affect a single coupling channel. 

In this work, we present a scheme for flexible quantum correlation manipulation based on a hybrid optomagnomechanical system comprising one atom mode, two orthogonally polarized optical modes, one magnon mode, and one phonon mode. Specifically, by tuning the polarization direction of the driving laser, different quantum correlation scenarios can be achieved on demand. Our all-optical controlled scheme features convenient operation, a simple setup, excellent feasibility, and high precision, as it simultaneously affects multiple coupling channels (optomechanical and effective Tavis-Cummings (TC) couplings).

\section{The hybrid optomagnomechanical system}
\begin{figure}[htbp]
\centering\includegraphics[width=8.0cm]{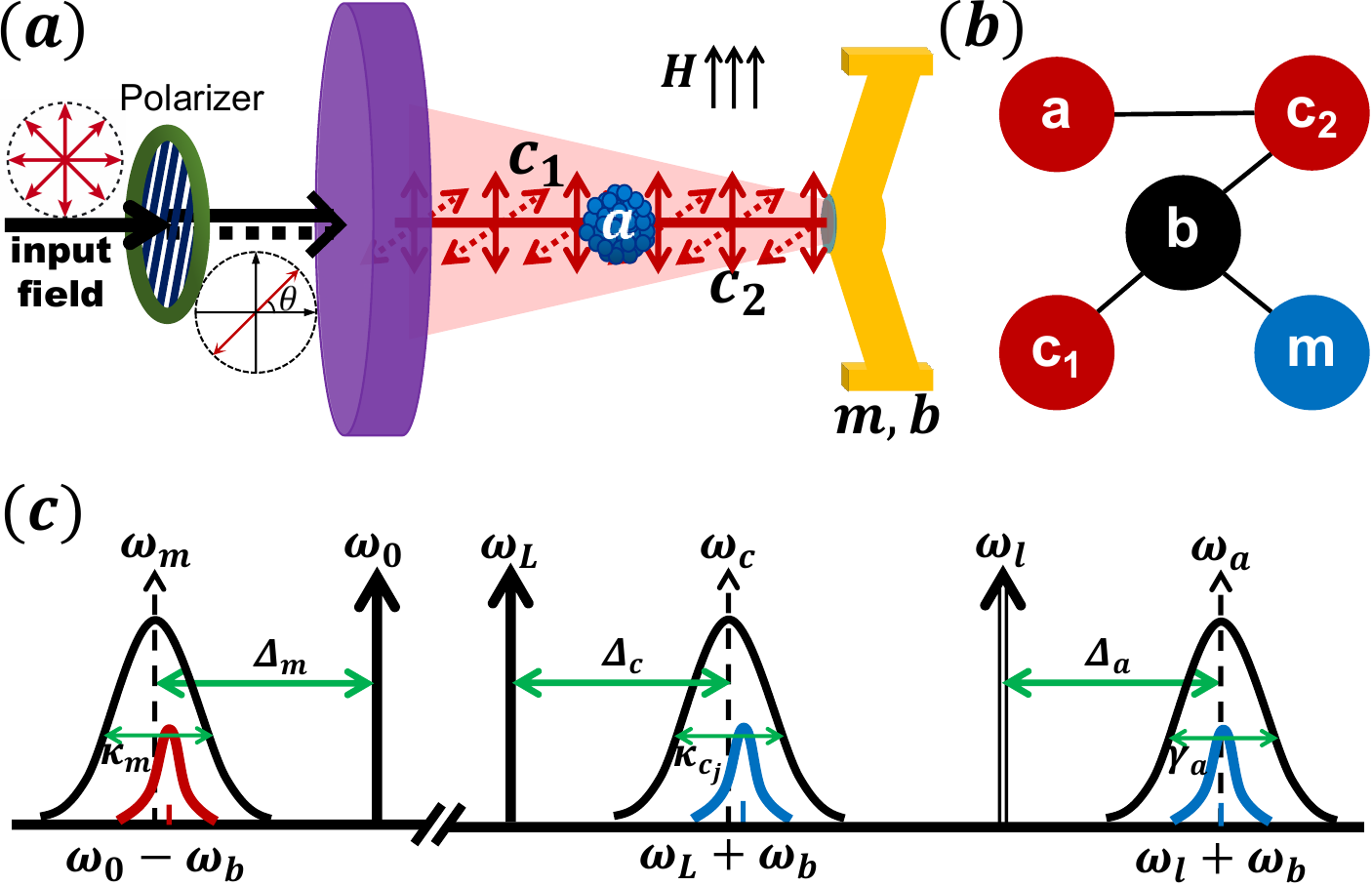}
\caption{The hybrid optomagnomechanical system. (a) Schematic diagram; (b) Interaction relation (blue/red circles denote blue/red-detuned driving); (c) Frequency relation.}
\end{figure}
Figure~1(a) depicts a hybrid optomagnomechanical system driven by linearly polarized light, whose polarization direction is tunable via rotating a polarizer. The system consists of an optical cavity formed by a mirror and a YIG microbridge~\cite{054031,015014} (its magnon (phonon) mode $m$ ($b$) has frequency $\omega_{m}$ ($\omega_b$) and decay rate $\kappa_m$ ($\gamma_b$)) and an ensemble of $N_a$ two-level atoms (with frequency $\omega_a$ and decay rate $\gamma_a$). The cavity mode can be decomposed into horizontal ($c_1$) and vertical ($c_2$) polarized modes at frequency $\omega_{c}$, with decay rates $\kappa_{c_1}$, $\kappa_{c_2}$, respectively. The interaction between the atomic ensemble and the cavity mode is a typical TC coupling. Notably, only the $c_2$ mode can effectively interact with the atoms, as its polarization direction is parallel to the static magnetic field orientation. Therefore, the corresponding coupling strength can be manipulated by adjusting the intensity weight ($|\sin\theta|$, with $\theta$ being the angle between the incident polarization and horizontal directions) of the $c_2$ mode, i.e., by rotating the polarizer.

To simplify the treatment of the system dynamics, the low-excitation limit for the atomic ensemble~\cite{050307,379} is considered. In other words, we assume that the atoms are initially prepared in the ground state and the excitation probability of a single atom is small, which is guaranteed by the off-resonant TC coupling. The collective spin operators of the atomic ensemble can be defined as $S_{\pm,z}=\sum_{i=1}^{N_a} \sigma_{\pm,z}^{(i)}$, where $\sigma_{\pm,z}$ denote the Pauli matrices, which obey the canonical commutation relations $[S_+, S_-]=S_z$ and $[S_z, S_{\pm}]=\pm 2 S_{\pm}$. Under this low-excitation approximation, $S_z\approx\langle S_z \rangle \approx -N_a$, and the dynamics of atomic polarization can be effectively described using bosonic operators. Specifically, the atomic annihilation operator is defined as $a=S_{-}/\sqrt{|S_z|}$, which satisfies the standard bosonic commutation relation $[\hat{a}, \hat{a}^\dagger] = 1$. 

Thus, the fully bosonized Hamiltonian of this system in the frame rotating at the driven fields with frequencies $\omega_L$, $\omega_l$ and $\omega_0$ can be described as:
\begin{equation} \begin{aligned}
 {\hat{H}}/{\hbar} &= \Delta_a \hat{a}^{\dagger} \hat{a}+\sum_{i=1}^2\Delta^0_{c_i} \hat{c}_i^{\dagger} \hat{c}_i+\Delta^0_m \hat{m}^{\dagger} \hat{m}+\frac{\omega_b}2{(\hat{q}^2+\hat{p}^2)}\\& + g_{m} \hat{m}^{\dagger}\hat{m}\hat{q} -\sum_{i=1}^2g_{c} \hat{c}_i^{\dagger}\hat{c}_i\hat{q}+g_{ac_2}|\sin\theta| (\hat{a}^{\dagger}\hat{c}_2+\hat{a}\hat{c}_2^{\dagger}) \quad \\&+i\eta_c|\cos{\theta}|(\hat{c}_1^{\dagger}- \hat{c}_1) +i\eta_c|\sin{\theta}|(\hat{c}_2^{\dagger}- \hat{c}_2)\\&+i\Omega_a(\hat{a}^{\dagger}- \hat{a})+i\Omega_m(\hat{m}^{\dagger}- \hat{m}), 
\end{aligned}\end{equation} 
where $\hat{j}^{\dagger}$ ($\hat{j}$), $j=a,c_1,c_2,m$, is the creation (annihilation) operator of the mode $j$. $\hat{p}$ and $\hat{q}$ are the dimensionless momentum and position operators of the phonon mode, respectively. $\Delta_{a}=\omega_{a}-\omega_{l}$, $\Delta_{c}^0=\omega_{c}-\omega_{L}$, and $\Delta_{m}^0=\omega_{m}-\omega_{0}$, are the frequency detunings of the atom, optical, and magnon modes for their driven fields, respectively. $\eta_c=\sqrt{2P\kappa_c/\hbar\omega_{c}}$ denotes the coupling strength between the input field and the optical modes, where $P$ is the input power. $\Omega_a=\frac{|d\cdot E_l|}{\hbar}\ll\omega_b$ denotes the Rabi frequency related to the laser drive (with amplitude $E_l$ and frequency $\omega_l$) and the atoms, where $d$ is the atomic dipole moment. $\Omega_m=\frac{\sqrt{5}}{4}\gamma\sqrt{N_d}B_d$ denotes the Rabi frequency related to the microwave drive (with amplitude $B_d$ and frequency $\omega_0$) and the magnon mode, where $\gamma$ is the gyromagnetic ratio and $N_d$ is the total number of spins of the YIG microbridge. The four terms in the first row of Eq.~(1) are the free Hamiltonians of the atom, optical, magnon, and phonon modes, respectively. The three terms in the second row describe the magnomechanical, optomechanical, and effective TC couplings, with coupling strengths $g_m$, $g_c$, and $g_{ac_2}|\sin\theta|$, respectively. The last two rows represent the driving terms of the optical, atom, and magnon modes, respectively.

From the Hamiltonian described by Eq.(1), the linearized quantum Langevin equations of the quadrature fluctuations ($\delta X_a$, $\delta Y_a$, $\delta X_{c_1}$, $\delta Y_{c_1}$, $\delta X_{c_2}$, $\delta Y_{c_2}$, $\delta X_{m}$, $\delta Y_{m}$, $\delta q$, $\delta p$ ) can be obtained as $\dot{u}=\mathcal{A} u + n(t)$ (details can be found in Appendix), where $\delta X_{j}=(\delta j+\delta j^{\dagger})/\sqrt{2}$ and $\delta Y_{j}=i(\delta j^{\dagger}-\delta j)/\sqrt{2}$, $j=a, c_1, c_2, m$. Then the corresponding covariance matrix (CM) can be obtained by solving the Lyapunov equation. Once the covariance matrix $V$ is obtained, bipartite entanglement, genuine tripartite entanglement, and quantum steering of the system can be quantitative measures by logarithmic negativity $E_N$~\cite{032314,090503}, minimum residual contangle $R^{min}$~\cite{7821,200615}, and steerability $S_{i\rightarrow j}$~\cite{060403}, respectively. 

\section{Generation of atom-magnon and optomagnonic entanglements}
\begin{figure}[htbp]
\centering\includegraphics[width=7cm]{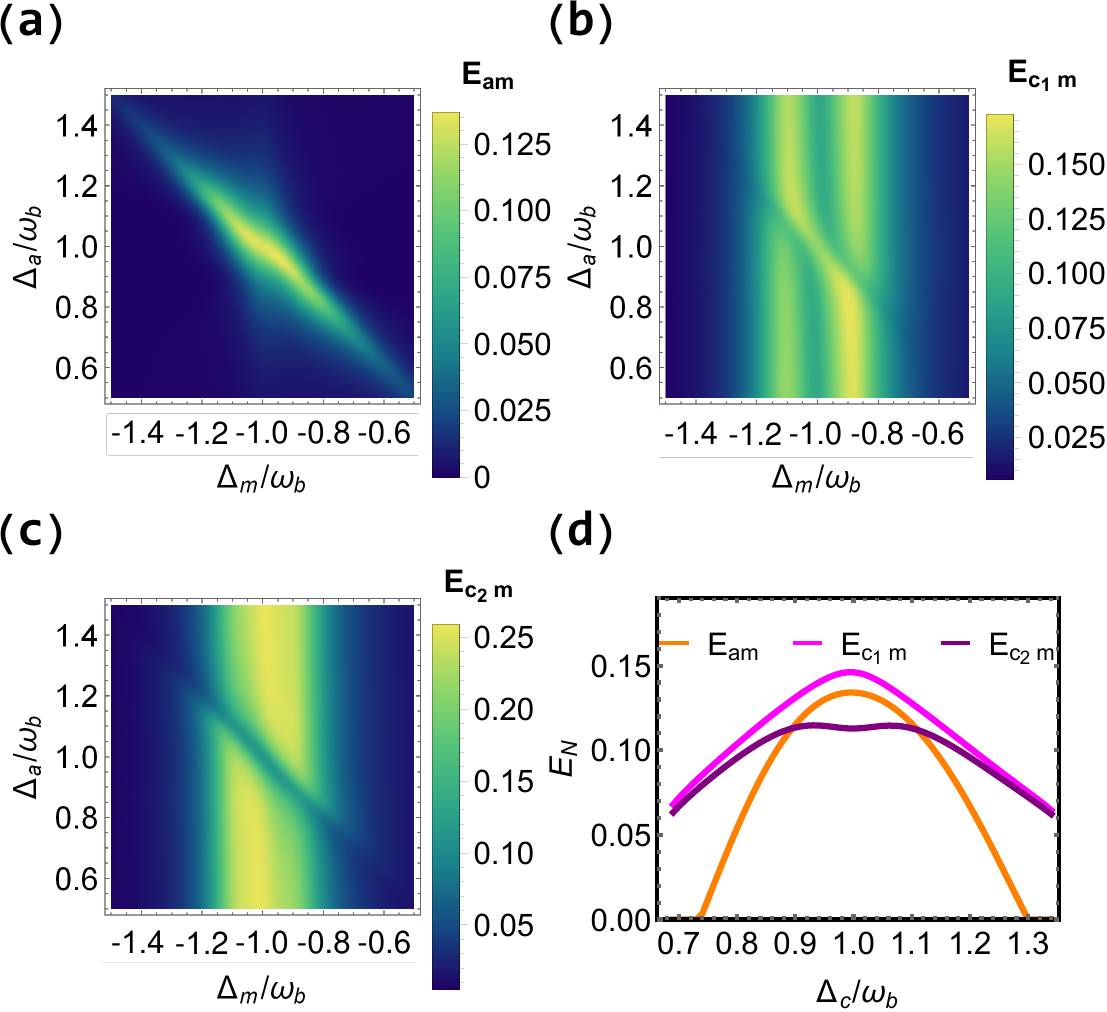}
\caption{Results of bipartite entanglements versus detunings: the atom-magnon entanglement $E_{am}$ (a), the optomagnonic entanglements $E_{c_1m}$ (b) and $E_{c_2m}$ (c) versus $\Delta_{a}$ and $\Delta_{m}$, and ($E_{am}$, $E_{c_1m}$, $E_{c_2m}$) versus $\Delta_{c}$ (d). Except for the varied parameters, we set $\omega_m/2\pi=10$~GHz, $\omega_b/2\pi=40$~MHz, $\gamma_a/2\pi=1$~MHz, $\kappa_{c_1}/2\pi=3$~MHz, $\kappa_{c_2}/2\pi=1$~MHz, $\kappa_m/2\pi=1$~MHz, $\gamma_b/2\pi=100$~Hz, $g_{ac_2}/2\pi=3$~MHz, $\theta=\pi/4$, $T=10$~mK, $G_c/2\pi=10$~MHz, $G_m/2\pi=2$~MHz, and $\Delta_a=-\Delta_m=\Delta_{c}=\omega_b$.}
\end{figure}
The magnon-phonon entanglement is created via magnomechanical coupling (with the magnon mode $m$ driven by blue-detuning), then transferred to the optical modes $c_1$ and $c_2$ (driven by red-detuning) via optomechanical coupling, and subsequently routed further to the atom mode $a$ through TC coupling, as illustrated in Fig.~1(b). The corresponding frequency relation is shown in Fig.~1(c). Accordingly, atom-magnon entanglement $E_{am}$ and optomagnonic entanglements $E_{c_1m}$ and $E_{c_2m}$ can be obtained; these three entanglements as functions of the effective magnon detuning $\Delta_{m}$ and the atom detuning $\Delta_{a}$ are presented in Figs.~2(a)-(c), respectively. Here $\Delta_{j}=\Delta_{j}^{0}-g_{j} q_{s}$ ($j=c,m$) and $q_s$ is the steady-state solution of the phonon mode. Clearly, all three entanglements can be maximized near $\Delta_m=-\omega_b$ (the optimal condition for entanglement generation); the two optical modes $c_1$ and $c_2$ compete with each other, resulting in complementarity between $E_{c_2m}$ and $E_{c_1m}$, while a portion of $E_{c_2m}$ is transferred to $E_{am}$. The variation of these three entanglements with the effective optical detuning is depicted in Fig.~2(d), and all are maximized near $\Delta_{c}=\omega_b$ (the optimal condition for state transfer).

\begin{figure}[htbp]
\centering\includegraphics[width=\linewidth]{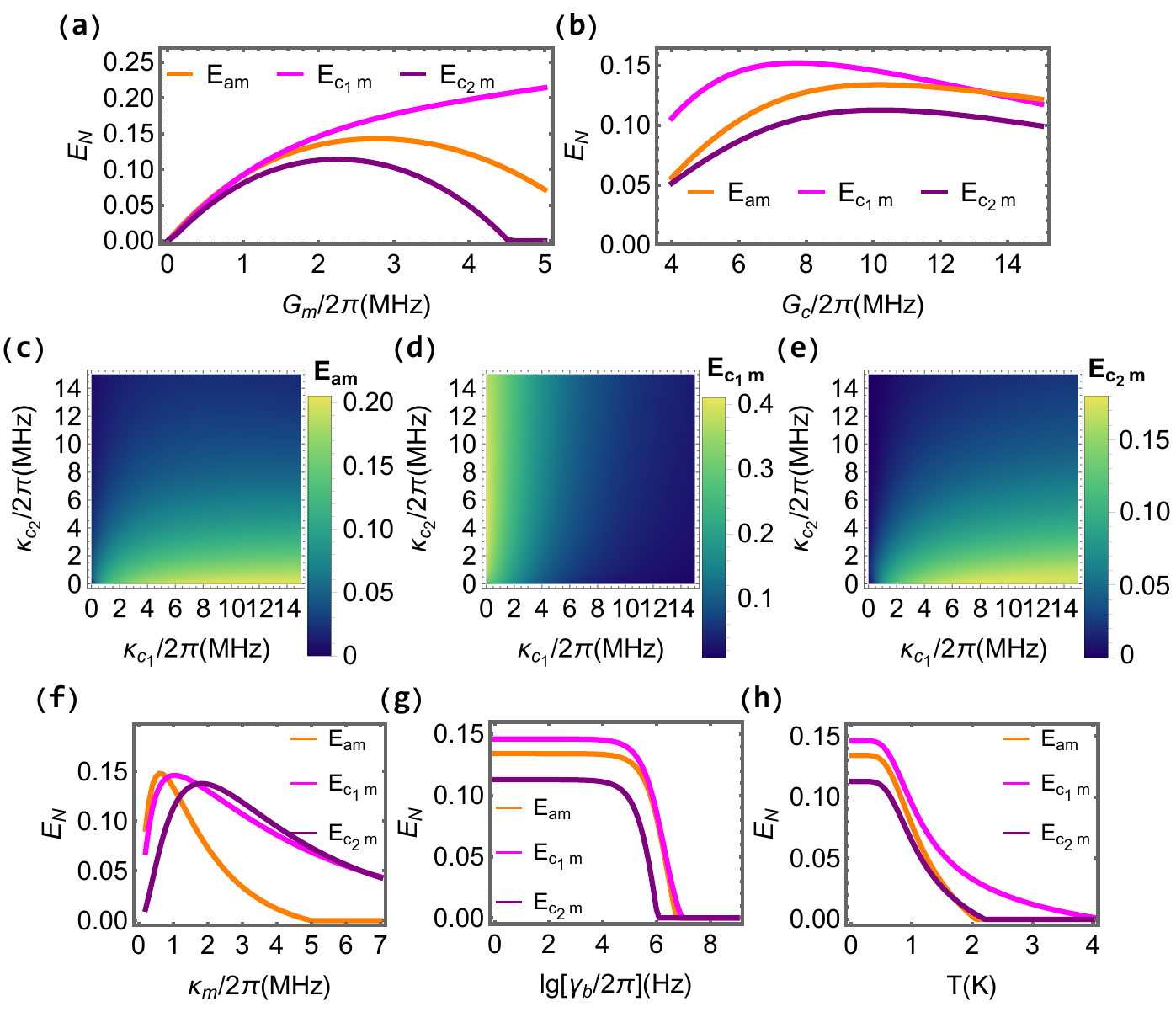}
\caption{Results of bipartite entanglements versus coupling and decoherence factors: ($E_{am}$, $E_{c_1m}$, $E_{c_2m}$) versus the effective magnomechanical coupling strength $G_{m}$ (a), the effective optomechanical coupling strength $G_{c}$ (b), the magnon decay rate $\kappa_{m}$ (f), the phonon decay rate $\gamma_b$ (g), and the temperature $T$ (h); $E_{am}$ (c), $E_{c_1m}$ (d), $E_{c_2m}$ (e) versus the optical decay rates $\kappa_{c_1}$ and $\kappa_{c_2}$. All parameters remain the same as those in Fig.~2.}
\end{figure}
The dependence of these three entanglements on effective magnomechanical and optomechanical coupling strengths $G_{m}$ and $G_c$ is shown in Figs.~3(a) and 3(b), respectively, where $G_{j}=\sqrt{2}g_{j} j_{s}$ ($j=c,m$). Here $c_{s}$ and $m_s$ are the steady-state solutions of the total optical and magnon modes (see Appendix for details). It is found that the entanglements exhibit a \enquote{first increasing and then decreasing} trend with increasing coupling strengths. This is because an increase in coupling strength initially exerts a positive effect on the generation or transfer of entanglement; however, when the coupling strength increases to a certain extent, it becomes detrimental due to the mutual constraints between the entanglement generation and transfer mechanisms. 
The robustness of entanglement against decoherence is a central concern for practical applications. Therefore, the effect of different factors that may introduce decoherence, including various decay rates and environmental temperature, is examined in Figs.~3(c)-(h). It is found from Figs.~3(c)-(e) that for the optical mode $c_1$ ($c_2$), increasing its own decay rate $\kappa_{c_1}$ ($\kappa_{c_2}$) degrades its entanglement transfer capability, resulting in a decrease in the self-related entanglement $E_{c_1m}$ ($E_{c_2m}$) and an increase in the entanglement associated with its competing counterpart $E_{c_2m}$ ($E_{c_1m}$). Due to their common entanglement source, the dependencies of $E_{am}$ and $E_{c_2m}$ on the optical decay rates are coincident. It can be observed from Fig.~3(f) that the entanglements exhibit a brief increase followed by a gradual decline as the magnon decay rate $\kappa_m$ increases. This behavior arises because dissipative coupling facilitates entanglement generation at small $\kappa_m$; however, when the decay rate becomes sufficiently large, dissipation-induced decoherence dominates, resulting in the degradation of the entanglements. 
The influence of the phonon decay rate and the environmental temperature on the entanglements is shown in Figs.~3(g) and 3(h), respectively. The entanglements remain steady under the conditions that $\gamma_b<2\pi\times100$~kHz ($T=10$~mK) or $T<0.4$~K ($\gamma_b=2\pi\times100$~Hz). Notably, $E_{am}$ and $E_{c_2m}$ can exist for $T<2$~K, and $E_{c_1m}$ persist until 4~K, confirming the considerable robustness of the generated quantum correlations against thermal noise. 

\section{Manipulation of entanglement and steering}
\begin{figure}[htbp]
\centering\includegraphics[width=\linewidth]{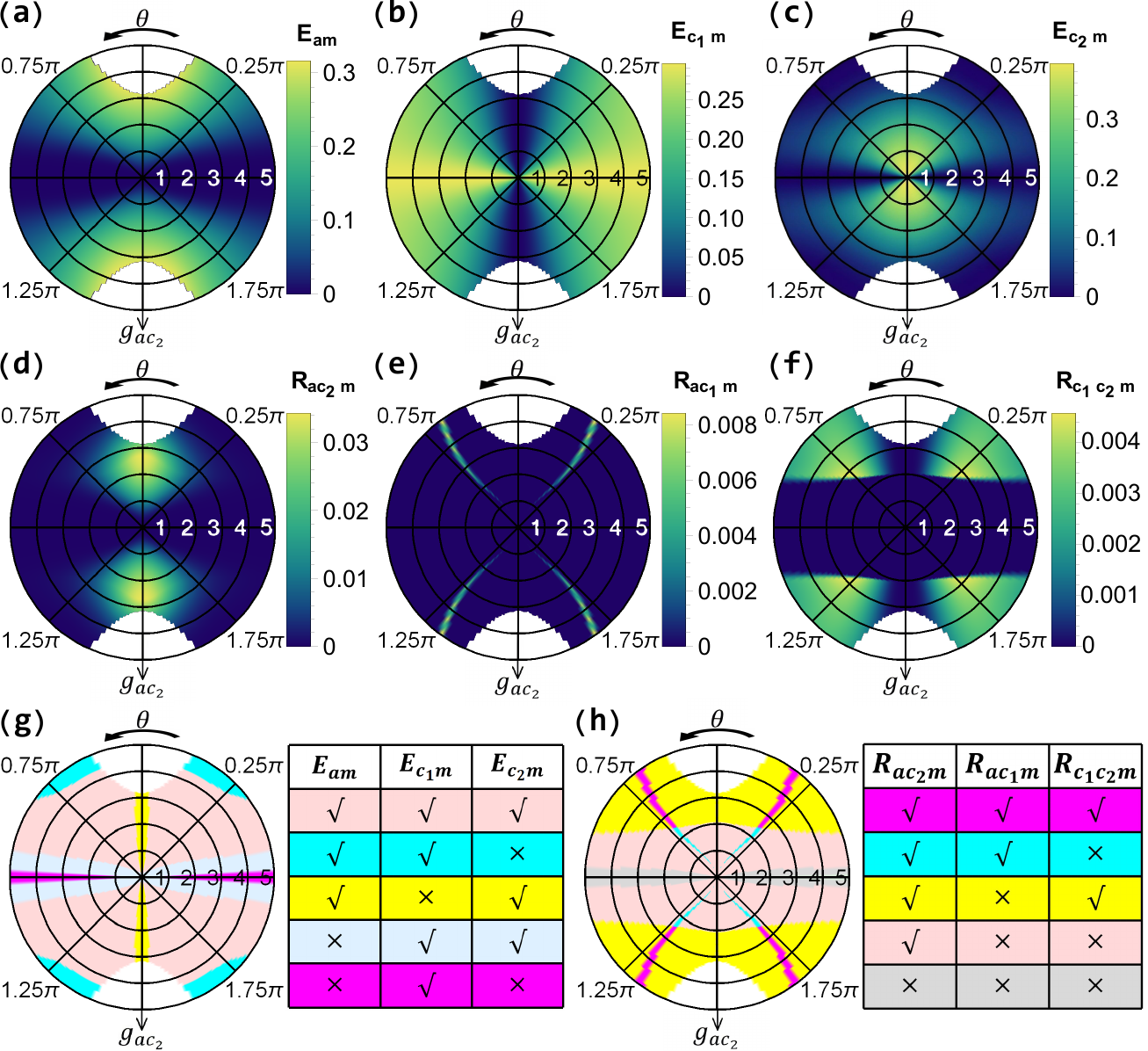}
\caption{Results of bipartite and genuine tripartite entanglements versus the TC coupling strength $g_{ac_2}$ and the polarization direction $\theta$: $E_{am}$ (a), $E_{c_1m}$ (b), $E_{c_2m}$ (c), $R_{ac_2m}$ (d), $R_{ac_1m}$ (e), and $R_{c_1c_2m}$ (f). The results of bipartite and genuine tripartite entanglements are summarized in (g) and (h), respectively. We set $G_m/2\pi=3$~MHz, and all other parameters remain the same as those in Fig.~2.}
\end{figure}
To match different application situations, it is necessary to explore the manipulation of the generated quantum correlations. Since the polarization direction $\theta$ and the TC coupling strength $g_{ac_2}$ can be conveniently adjusted in practical experiments, various quantum correlations as functions of $\theta$ and $g_{ac_2}$ are examined in this section. The optomagnonic and atom-magnon entanglements versus them are depicted in Figs.~4(a)-(c). It is found that for $E_{am}$ and $E_{c_2m}$ ($E_{c_1m}$), the strongest results appear near $\theta=\pi/2,3\pi/2$ ($\theta=0,\pi$) where the optical mode $c_2$ ($c_1$) dominates; the parameter area where entanglements exist becomes wider with increasing TC coupling strength $g_{ac_2}$; the entanglement transfer from $E_{c_2m}$ to $E_{am}$ and the complementarity between $E_{c_2m}$ and $E_{c_1m}$ are also clearly shown. These three entanglements are collectively presented in Fig.~4(g). Near $\theta=0$, only $E_{c_1m}$ exists; with the increase of $\theta$, $E_{c_2m}$ and $E_{am}$ appear in sequence; $E_{c_1m}$ disappears near $\theta=\pi/2$ and $E_{c_2m}$ disappears for large $g_{ac_2}$. The flexibility of the system in manipulating bipartite entanglement lies in: by tuning $\theta$ and $g_{ac_2}$, the magnon mode can be entangled with the optical $c_1$ mode, $c_2$ mode, or both; on this basis, whether the magnon mode is also entangled with the atom mode can be further determined. Notably, the system becomes unstable near $\theta=\pi/2,3\pi/2$ for large $g_{ac_2}$, as the competition of the atom mode for the $c_2$ mode impede effective phonon cooling. Various genuine tripartite entanglements versus $\theta$ and $g_{ac_2}$ are presented in Figs.~4(d)-(f). Clearly, $R_{ac_1m}$ and $R_{c_1c_2m}$ mainly exist for the balanced $c_1$ and $c_2$ modes (near $\theta=(2n+1)\pi/4$), while $R_{ac_2m}$ under $c_2$-dominated conditions (near $\theta=\pi/2,3\pi/2$). It seems that the optimal genuine tripartite entanglements require moderate TC coupling (not too small, not too large). These genuine tripartite entanglements are summarized in Fig.~4(h). The system exhibits flexible manipulation of genuine tripartite entanglement: by adjusting $\theta$ and $g_{ac_2}$, one can decide whether $R_{ac_2m}$ exist; when $R_{ac_2m}$ is present, it is further feasible to manipulate the existence of $R_{ac_1m}$, $R_{c_1c_2m}$, or both.

\begin{figure}[htbp]
\centering\includegraphics[width=\linewidth]{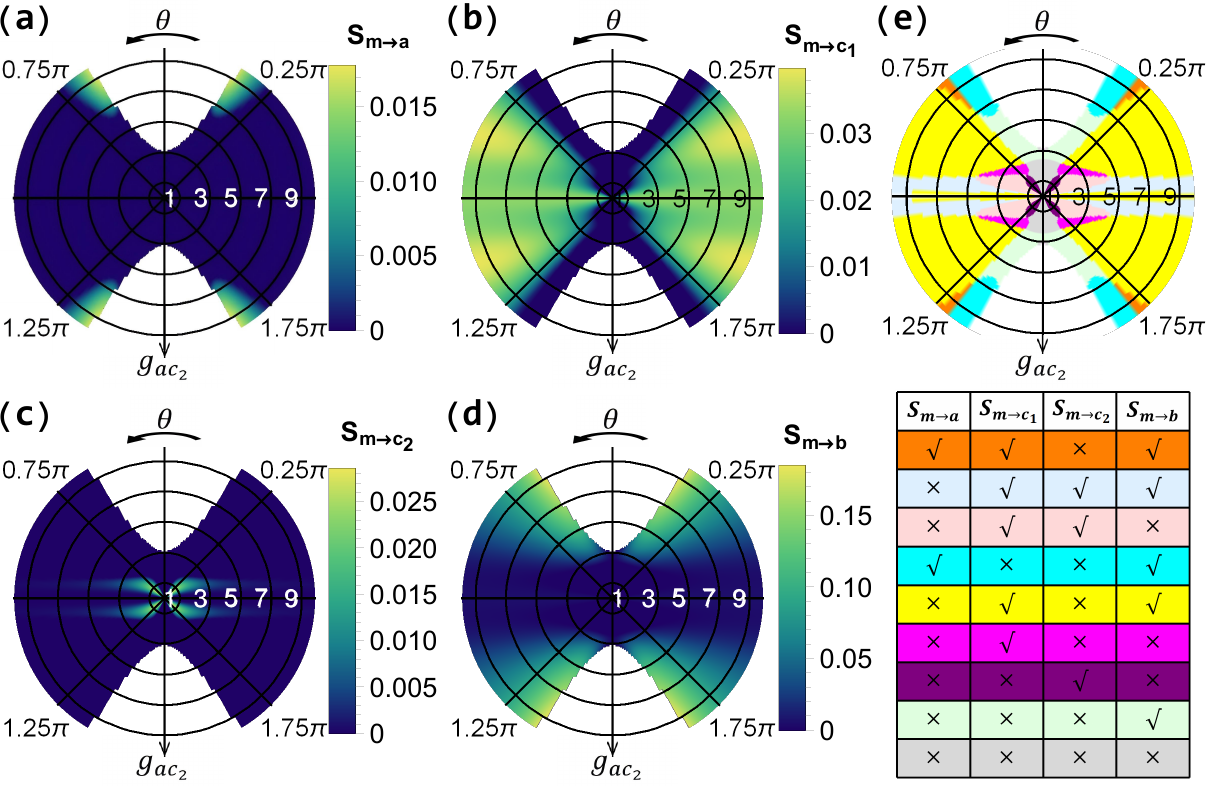}
\caption{Results of bipartite steerings versus $g_{ac_2}$ and $\theta$: $S_{m\rightarrow a}$ (a), $S_{m\rightarrow c_1}$ (b), $S_{m\rightarrow c_2}$ (c), and $S_{m\rightarrow b}$ (d). All results are summarized in (e). The blank area represents the unstable regime. All parameters remain the same as those in Fig.~4.}
\end{figure}
The possible bipartite quantum steerings versus $g_{ac_2}$ and $\theta$ are illustrated in Figs.~5(a)-(d). Only one-way quantum steerings with the magnon mode acting as the steering party can emerge due to its high occupancy~\cite{032121,022335}. The steerings from the magnon mode to the atom and phonon modes primarily emerge in the parameter regime where the optical $c_2$ mode dominates and the TC coupling is strong enough (with $S_{m\rightarrow a}$ requiring a larger $g_{ac_2}$), as illustrated in Figs.~5(a) and 5(d), respectively. Meanwhile, the steering from the magnon mode to the optical $c_1$ mode exists within the parameter regime dominated by the optical $c_1$ mode, as shown in Fig.~5(b). Specifically, the maximum steering can be achieved near $\theta=0$ for small $g_{ac_2}$. When $g_{ac_2}$ is large enough, the maximum steering emerges near $\theta=\pi/4$. As indicated in Fig.~5(c), the steering from the magnon mode to the optical $c_2$ mode $S_{m\rightarrow c_2}$ can be maximized for almost balanced $c_1$ and $c_2$ modes when $g_{ac_2}$ is weak. For large $g_{ac_2}$, $S_{m\rightarrow c_2}$ is weak and gradually shifted to the $c_1$-dominated regime. The maximum of $S_{m\rightarrow c_1}$ and $S_{m\rightarrow c_2}$ presented in Figs.~5(b) and 5(c) can be understood by considering both the existence and optimization conditions of quantum steering. Provided quantum steering exists, the optimal steerability can be achieved by minimizing the relative occupancy of the steered party. Additionally, the existence regions for various combinations of the above four one-way steerings are illustrated in Fig.~5(e). As can be seen, tuning $g_{ac_2}$ and $\theta$ yields four scenarios: the absence of one-way steering, the presence of single one-way steering ($S_{m\rightarrow c_1}$, $S_{m\rightarrow c_2}$, or $S_{m\rightarrow b}$), the coexistence of two one-way steerings ($S_{m\rightarrow c_1}$ \& $S_{m\rightarrow c_2}$, $S_{m\rightarrow c_1}$ \& $S_{m\rightarrow b}$, $S_{m\rightarrow a}$ \& $S_{m\rightarrow b}$), and the coexistence of three one-way steerings ($S_{m\rightarrow c_1}$ \& $S_{m\rightarrow c_2}$ \& $S_{m\rightarrow b}$, $S_{m\rightarrow c_1}$ \& $S_{m\rightarrow a}$ \& $S_{m\rightarrow b}$). Notably, $S_{m\rightarrow a}$ cannot coexist with $S_{m\rightarrow c_2}$ due to correlation transfer induced by TC coupling~\cite{022332}. Such flexible manipulation of quantum steering holds potential for various quantum communication tasks, enabling dynamic adjustments to the role and number of the steered party. 

\begin{figure}[htbp]
\centering\includegraphics[width=\linewidth]{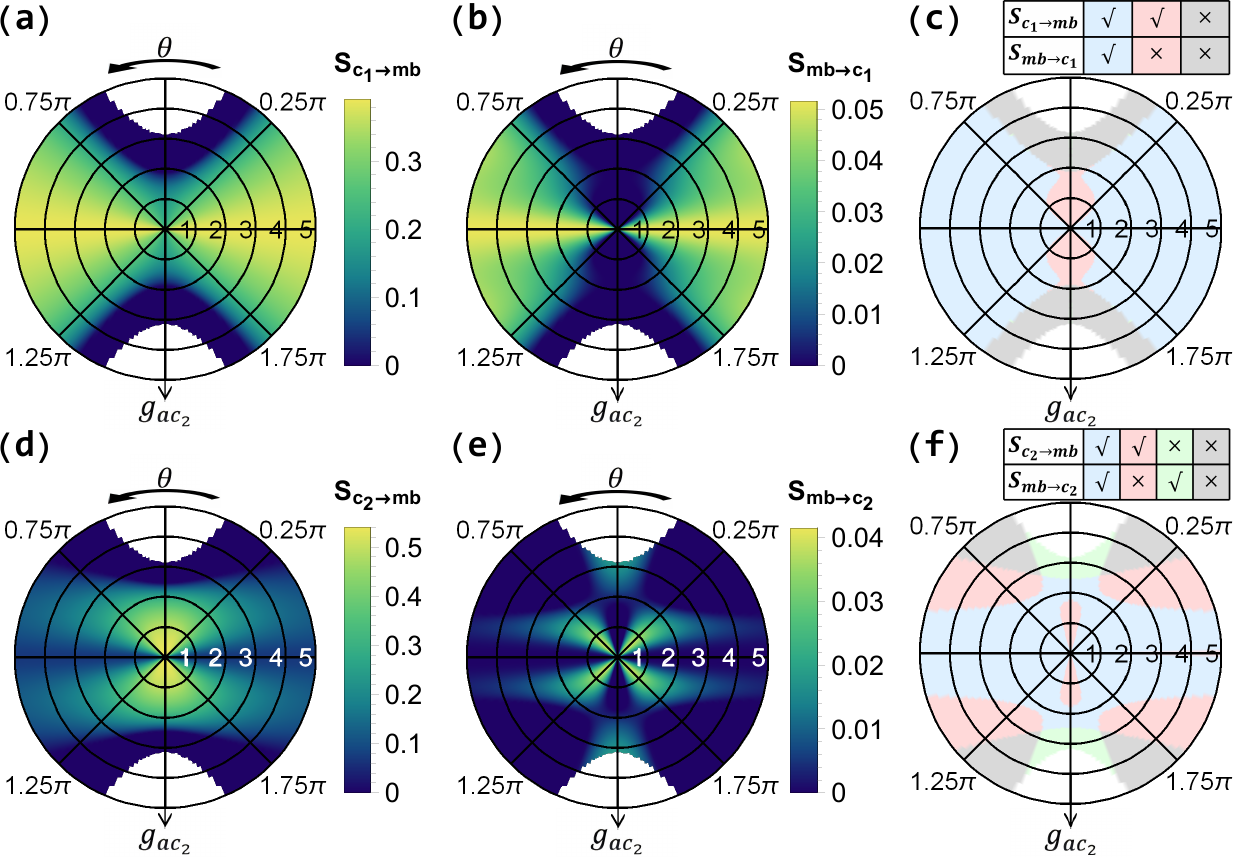}
\caption{Results of tripartite steerings versus $g_{ac_2}$ and $\theta$: $S_{c_1\rightarrow mb}$ (a), $S_{mb\rightarrow c_1}$ (b), $S_{c_2\rightarrow mb}$ (d) and $S_{mb\rightarrow c_2}$ (e). The results involving $c_1$ and $c_2$ are summarized in (c) and (f), respectively. The blank area represents the unstable regime. All parameters remain the same as those in Fig.~4.}
\end{figure}
The tripartite quantum steerings between the optical mode $c_1$ ($c_2$) and the magnon-phonon joint mode $mb$ as functions of $g_{ac_2}$ and $\theta$ are illustrated in Figs.~6(a)-(b) (Figs.~6(d)-(e)) and summarized in Fig.~6(c) (Fig.~6(f)). As shown in Figs.~6(a)-(b), $c_1$ cannot steer individual $m$, whereas $S_{c_1\rightarrow mb}$ is strong and persists over a wide parameter region. Meanwhile, the joint steering $S_{mb\rightarrow c_1}$ is stronger than $S_{m\rightarrow c_1}$. These behaviors can be understood in terms of the type-III monogamy relation. By tuning $g_{ac_2}$ and $\theta$, one can realize no-way steering ($S_{c_1\rightarrow mb}=0$ \& $S_{mb\rightarrow c_1}=0$, gray region), one-way steering ($S_{c_1\rightarrow mb}>0$ \& $S_{mb\rightarrow c_1}=0$, pink region), and two-way steering ($S_{c_1\rightarrow mb}>0$ \& $S_{mb\rightarrow c_1}>0$, blue region) on demand, as illustrated in Fig.~6(c). Corresponding results for the optical mode switched from $c_1$ to $c_2$ are displayed in Figs.~6(d)–(e). Notably, the reverse one-way steering ($S_{c_2\rightarrow mb}=0$ \& $S_{mb\rightarrow c_2}>0$, green area) becomes accessible; therefore, the manipulation of the direction of the one-way steering is available. 

\begin{figure}[htbp]
\centering\includegraphics[width=\linewidth]{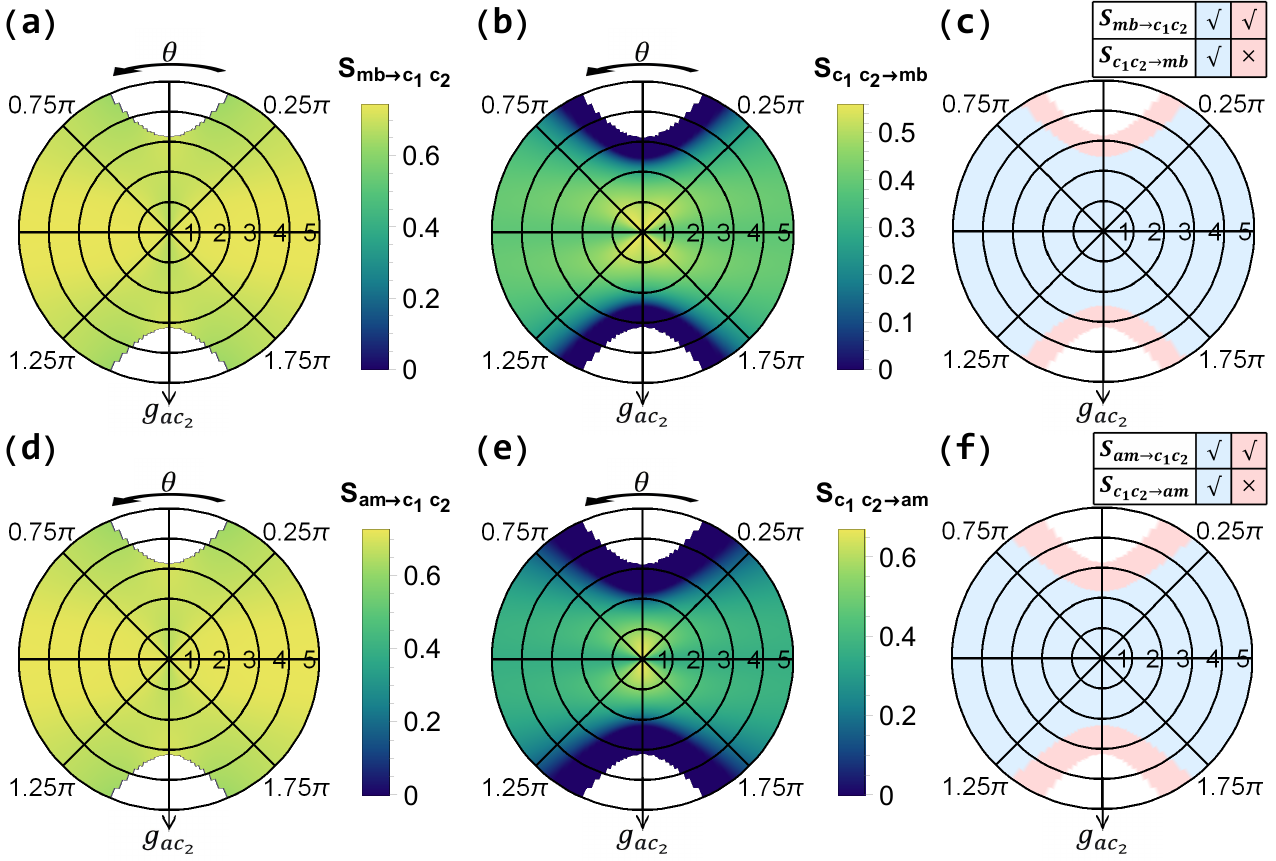}
\caption{Results of quadripartite steerings versus $g_{ac_2}$ and $\theta$: $S_{mb\rightarrow c_1c_2}$ (a), $S_{c_1c_2\rightarrow mb}$ (b), $S_{am\rightarrow c_1c_2}$ (d), $S_{c_1c_2\rightarrow am}$ (e). The results involving $mb$ and $am$ are summarized in (c) and (f), respectively. The blank area represents the unstable regime. All parameters remain the same as those in Fig.~4.}
\end{figure}
The (2+2) quadripartite steerings between the joint optical mode $c_1c_2$ and the joint magnon-phonon (atom-magnon) mode $mb$ ($am$), as functions of $g_{ac_2}$ and $\theta$, are illustrated in Figs.~7(a)-(b) (Figs.~7(d)-(e)) and summarized in Fig.~7(c) (Fig.~7(f)). As shown in Figs. 7(a)–(b) and 7(d)–(e), the steerability becomes stronger and persists over a wider parameter region when the joint optical mode $c_1c_2$ serves as the steered party. By tuning $g_{ac_2}$ and $\theta$, one can realize on demand two types of quadripartite steering: one-way steering (stationary qubit $mb$ or $am$$\rightarrow $flying qubit $c_1c_2$, pink region), and two-way steering (stationary qubit $mb$ or $am$$\leftrightarrow $flying qubit $c_1c_2$, blue region), as illustrated in Fig.~7(c) and Fig.~7(f). 
\begin{figure}[htbp]
\centering\includegraphics[width=4.5cm]{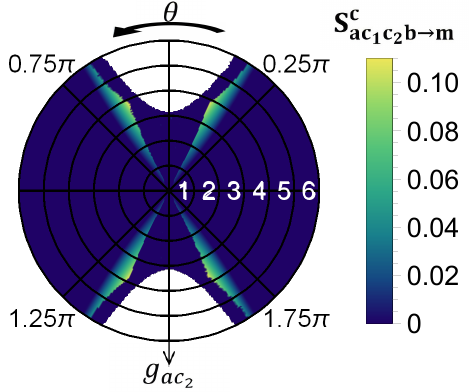}
\caption{The collective pentapartite steering $S_{ac_1c_2b\rightarrow m}^c$ versus $g_{ac_2}$ and $\theta$. The blank area represents the unstable regime. All parameters remain the same as those in Fig.~4.}
\end{figure}
Furthermore, the collective pentapartite steering $S_{ac_1c_2b\rightarrow m}^c$ is also investigated, as shown in Fig.~8. Within four special parameter regions which are symmetrically distributed, $S_{ac_1c_2b\rightarrow m}>0$ and $S_{ac_1c_2\rightarrow m}=S_{c_1c_2b\rightarrow m}=S_{ac_1b\rightarrow m}=S_{ac_2b\rightarrow m}=0$, i.e., the magnon mode can be steered only by jointing all other modes. Our results suggest that choosing a moderate $g_{ac_2}$ and an appropriate $\theta$ within the $c_2$-dominated parameter region may facilitate the realization of this collective steering, which provides enhanced security in multipartite quantum protocols, as it prevents unauthorized access by any subset of the steering party. 

Overall, coordinated tuning of the polarization direction $\theta$ and the TC coupling strength $g_{ac_2}$ enables a effective manipulation over diverse quantum correlations—including bipartite and genuine tripartite entanglements, bipartite and multipartite steerings, as well as collective pentapartite steering. This all-optical control scheme provides a versatile and experimentally feasible approach to dynamically configuring quantum resources in integrated hybrid systems.

\section{Conclusion}
In summary, this work proposes a scheme for generating and manipulating quantum correlations based on a hybrid optomagnomechanical system. The parameter dependence of the atom-magnon and optomagnonic entanglements is investigated, revealing robustness to external disturbance. Moreover, the selective generation and periodic manipulation of bipartite and genuine tripartite entanglements, as well as bipartite and multipartite steerings (including collective pentapartite steering), are achieved through precise tuning of both the polarization direction and the TC coupling strength. Notably, here multiple coupling channels can be tuned simultaneously. Therefore, our scheme enables the all-optical manipulation of the quantum correlations and offers high experimental feasibility toward integrated quantum information processing. This work may have potential applications in some special quantum tasks, such as quantum network construction, quantum information memory, ultra-safe multi-user quantum communication, optical control of magnons, etc. 

\section{Appendix}
The quantum Langevin equations (QLEs) that describe the system can be written as:
\begin{equation}\begin{aligned}
\dot{a} = & -(i\Delta_a+\gamma_a)a -ig_{ac_2}|\sin\theta| c_2 +\Omega_a+\sqrt{2\gamma_a}a_{in}\\
\dot{c}_1 = & -(i\Delta_{c}^{0}+\kappa_{c_1})c_1 +ig_{c}c_1q +\eta_c|\cos{\theta}|+\sqrt{2\kappa_{c_1}}c_{in1}\\
\dot{c}_2 = & -(i\Delta_{c}^{0}+\kappa_{c_2})c_2 -ig_{ac_2}|\sin\theta| a +ig_{c}c_2q +\eta_c|\sin{\theta}| \\&+\sqrt{2\kappa_{c_2}}c_{in2}\\
\dot{m} = & -(i\Delta_m^{0}+\kappa_m)m -ig_{m}mq +\Omega_m +\sqrt{2\kappa_m}m_{in}\\
\dot{q} = & \omega_b p\\
\dot{p} = & -\omega_b q-\gamma_b p +g_{c} c_1^\dagger c_1 +g_{c} c_2^\dagger c_2 -g_{m} m^\dagger m +\xi,
\end{aligned}\end{equation}
When the hybrid optomagnomechanical system is driven strongly, each mode can be expressed as the sum of its mean value and fluctuation. Finally, linearized QLEs that describe the time evolution of the quadrature operators can be expressed in matrix form as:
\begin{equation}
\dot{u}(t)=\mathcal{A} u(t)+n(t)
\end{equation}
where $u(t)=(\delta X_a, \delta Y_a, \delta X_{c_1}, \delta Y_{c_1}, \delta X_{c_2}, \delta Y_{c_2}, \delta X_{m}, \delta Y_{m},\\ \delta q, \delta p)^T$ is the vector of the quadrature fluctuation operators of the atom mode, the cavity modes, the magnon mode and the phonon mode. $n(t)$ is the vector of noise quadrature operators associated with the noise terms, and $\mathcal{A}$ is the drift matrix (the system is stable if every eigenvalue of matrix $\mathcal{A}$ has a negative real part), which can be explicitly written as: 
\begin{equation}\mathcal{A}=\begin{pmatrix}
 \mathcal{A}_a & 0 & \mathcal{A}_{ac_2} & 0 & 0\\
 0 & \mathcal{A}_{c_1} & 0 & 0 & \mathcal{A}_{c_1b}\\
 \mathcal{A}_{ac_2}^* & 0 & \mathcal{A}_{c_2} & 0 & \mathcal{A}_{c_2b}\\
 0 & 0 & 0 & \mathcal{A}_m & \mathcal{A}_{mb}\\
 0 & \mathcal{A}_{c_1b}^* & \mathcal{A}_{c_2b}^* & \mathcal{A}_{mb}^* & \mathcal{A}_b\\
\end{pmatrix},\end{equation}
where each element denotes a $2\times2$ block matrix and the non-zero block matrices are $\mathcal{A}_a=\begin{pmatrix} -\gamma_a & \Delta_{a}\\ -\Delta_{a} &-\gamma_a\end{pmatrix}$, $\mathcal{A}_{ac_2}=\mathcal{A}_{ac_2}^*=\begin{pmatrix} 0 & g_{ac_2}|\sin\theta|\\ -g_{ac_2}|\sin\theta| & 0\end{pmatrix}$, $\mathcal{A}_{m}=\begin{pmatrix} -\kappa_{m} & \Delta_{m}\\ -\Delta_{m} &-\kappa_{m}\end{pmatrix}$, $\mathcal{A}_{mb}=\begin{pmatrix} G_{m} & 0\\ 0 & 0\end{pmatrix}$, $\mathcal{A}_{mb}^*=\begin{pmatrix} 0 & 0\\ 0 & -G_{m}\end{pmatrix}$, $\mathcal{A}_{c_1}=\begin{pmatrix} -\kappa_{c_1} & \Delta_{c_1}\\ -\Delta_{c_1} &-\kappa_{c_1}\end{pmatrix}$, $\mathcal{A}_{c_2}=\begin{pmatrix} -\kappa_{c_2} & \Delta_{c_2}\\ -\Delta_{c_2} &-\kappa_{c_2}\end{pmatrix}$, $\mathcal{A}_{c_1b}=\begin{pmatrix} -G_{c}\cos\theta & 0\\ 0 & 0\end{pmatrix}$, $\mathcal{A}_{c_2b}=\begin{pmatrix} -G_{c}\sin\theta & 0\\ 0 & 0\end{pmatrix}$, $\mathcal{A}_{c_1b}^*=\begin{pmatrix} 0 & 0\\ 0 & G_{c}\cos\theta\end{pmatrix}$, $\mathcal{A}_{c_2b}^*=\begin{pmatrix} 0 & 0\\ 0 & G_{c}\sin\theta\end{pmatrix}$, $\mathcal{A}_{b}=\begin{pmatrix} 0 & \omega_{b}\\ -\omega_{b} &-\gamma_{b}\end{pmatrix}$. \\
$\Delta_{j}=\Delta_{j}^{0}-g_{j} q_{s}$ and $G_{j}=\sqrt{2}g_{j} j_s$, ($j=c,m$) denote the effective frequency detunings and the effective magnomechanical (optomechanical) coupling strength, respectively. Herein, $q_s=(g_mm_s-g_cc_{s})/\omega_b$, $m_s=\frac{\Omega_m}{\kappa_m+i\Delta_m}$ and $c_s=\sqrt{c_{1s}^2+c_{2s}^2}$ are the steady-state solutions of the phonon, magnon, and total optical modes, where $c_{1s}=\frac{\eta_c |\cos{\theta}|}{\kappa_{c_1}+i\Delta_{c}}$, $c_{2s}=\frac{\eta_c|\sin{\theta}|(\gamma_a+i\Delta_a)-ig_{ac_2}\Omega_a}{g_{ac_2}^2+(\gamma_a+i\Delta_a)(\kappa_{c_2}+i\Delta_{c})}$. 

The steady-state covariance matrix $V$ $(t\rightarrow\infty)$ of the system quadratures, with its elements defined as $V_{ij} = \langle{\{u_i(t),u_j(t)\}}\rangle/2$, can be obtained by solving the Lyapunov equation:
\begin{equation}
\mathcal{A}V+V\mathcal{A}^{T}=-\mathcal{D}
\end{equation}
where $\mathcal{D}$ is the diffusion matrix, with its entries defined as
$\langle{n_i(t) n_j(t^{\prime}) +n_j(t^{\prime}) n_i(t)}\rangle/2=\mathcal{D}_{ij}\delta(t-t^{\prime})$. For our system,  $\mathcal{D}=\text{Diag}(\gamma_a, \gamma_a, \kappa_{c_1}, \kappa_{c_1}, \kappa_{c_2}, \kappa_{c_2}, \kappa_m(2{\overline{n}_m}+1), \kappa_m(2{\overline{n}_m}+1), 0, \gamma_{b}(2{\overline{n}_b}+1))$. $a_{in}(t)$, $c_{in1}(t)$, $c_{in2}(t)$, $m_{in}(t)$ and $\xi(t)$ are the noise operators of the atom mode $a$, the optical modes $c_1,c_2$, the magnon mode $m$, and the phonon mode $b$, respectively, which have zero means and are characterized by the following correlation function: 
\begin{equation}\begin{aligned}
\left \langle o_{in}(t)o_{in}^{\dagger}(t^{\prime}) \right \rangle&=\delta(t-t^{\prime}),\\
\left \langle m_{in}(t)m_{in}^{\dagger}(t^{\prime}) \right \rangle&=({\overline{n}_m+1})\delta(t-t^{\prime}),\\
\left \langle m_{in}^{\dagger}(t)m_{in}(t^{\prime}) \right \rangle&={\overline{n}_m}\delta(t-t^{\prime}),\\
\left \langle \xi(t)\xi(t^{\prime})+\xi(t^{\prime}) \xi(t)\right \rangle/2 &\simeq\gamma_{b}(2{\overline{n}_b}+1)\delta(t-t^{\prime}), \end{aligned}\end{equation}
where $o=a,c_1,c_2$, ${\overline{n}_m}=\left[\exp[(\hbar \omega_{m}/k_{B}T)]-1\right]^{-1}$ (${\overline{n}_b}=\left[\exp[(\hbar \omega_{b}/k_{B}T)]-1\right]^{-1}$) is the equilibrium mean magnon (phonon), and a Markovian approximation has been made.

The bipartite entanglement is quantified by the logarithmic negativity, defined as:
\begin{equation}
E_N=\max[0, -\ln2\nu_{-}]
\end{equation}
where $\nu_{-} = \min \thinspace eig\lvert i\Omega_2V_m\rvert$ ($\Omega_2= \oplus_{j=1}^2 i \sigma_y $ is the so-called symplectic matrix, and $\sigma_y$ is the Pauli $y$ matrix) is the minimum symplectic eigenvalue of the covariance matrix $V_m= PVP$, with $V_m$ being the $4 \times 4$ covariance matrix associated, and $P = \text{Diag}\begin{pmatrix} 1, 1, 1, -1 \end{pmatrix}$, realizing partial transposition at the level of covariance matrices. $E_N>0$ denotes the presence of bipartite entanglement in the system.

The genuine tripartite entanglement is quantified by the minimum residual contangle, defined as:
\begin{equation}
R_{min}=\min[R_{i|jk},R_{j|ik},R_{k|ij}]
\end{equation}
where $R_{i|jk}=C_{i|jk}-C_{i|j}-C_{i|k}\geq 0$ is the residual contangle, with $C_{u|v}=E_{u|v}^2$ being the contangle of subsystems $u$ and $v$ ($v$ contains one or two modes), which is a proper entanglement monotone defined as squared logarithmic negativity. $C_{i|jk}=\max[0, -\ln2\nu_{-}^\prime]^2, \nu_{-}^\prime=\min \thinspace eig\lvert i\Omega_3V_{m^\prime}\rvert$ is the minimum symplectic eigenvalue of the covariance matrix $V_{m^\prime}=P_{i|jk}V^\prime P_{i|jk}$, where $V^\prime$ is the $6\times 6$ covariance matrix of the whole system, $\Omega_3= \oplus_{j=1}^3 i\sigma_y$ and $P_{i|jk} = \text{Diag}\begin{pmatrix}1, -1, 1, 1, 1, 1\end{pmatrix}$. For $R_{j|ik}$ and $R_{k|ij}$, the calculation process is similar to $R_{i|jk}$, with $P_{j|ik} = \text{Diag}\begin{pmatrix}1, 1, 1, -1, 1, 1\end{pmatrix}$ and $P_{k|ij} = \text{Diag}\begin{pmatrix}1, 1, 1, 1, 1, -1\end{pmatrix}$. $R^{min}>0$ denotes the presence of genuine tripartite entanglement in the system.

Moreover, the steering is characterized by the steerabilities between the two subsystems, which are:
\begin{equation}
\begin{aligned}
S_{1\rightarrow 2}=\max[0, \frac{1}{2}\ln\frac{\text{Det}(V_{11})}{4\text{Det}(V)}] \\
S_{2\rightarrow 1}=\max[0, \frac{1}{2}\ln\frac{\text{Det}(V_{22})}{4\text{Det}(V)}]
\end{aligned}
\end{equation}
where, $V=\begin{pmatrix}
 V_{11} & V_{12}\\
 V_{21} & V_{22}
\end{pmatrix}$, $V_{11}$ and $V_{22}$ are the two subsystems, and $V_{12}$ and $V_{21}$ are the associated items between subsystem-1 and 2. $S_{1\rightarrow 2}>0$ ($S_{2\rightarrow 1}>0$) indicates that subsystem-1 can steer subsystem-2 (subsystem-2 can steer subsystem-1). One-way steering emerges when $S_{1\rightarrow 2}>0$ \& $S_{2\rightarrow 1}=0$, or $S_{2\rightarrow 1}>0$ \& $S_{1\rightarrow 2}=0$; in contrast, two-way steering occurs under the condition $S_{1\rightarrow 2}>0$ \& $S_{2\rightarrow 1}>0$.

For a quantum system consisting of $n$ parties, a given party $i$ exhibits collective multipartite steering if it can be steered by all the remaining $n-1$ parties, yet cannot be steered by any subset of $n-2$ parties. Considering a $n$-partite case that has a steered mode ($B$) and $n-1$ steering modes ($A_1,A_2,...,A_{n-1}$), the criterion of the collective multipartite steering can be simplified as:
\begin{equation}
 \begin{aligned}
 S_{A_1A_2...A_{n-2}\rightarrow B}=0 \\
 S_{A_1A_2...A_{n-1}\rightarrow B}>0
 \end{aligned}
\end{equation}

This work was supported by the Innovation Program for Quantum Science and Technology (Grant No. 2023ZD0300400), the National Key Research and Development Program of China (Grant No. 2021YFC2201802), and the Central Government Guidance Funds for Local Science and Technology Development Projects (Grant No. YDZJSX2025D001).

\bibliography{doc/latex/revtex/aip/ref}
\end{document}